%% file: extended-abstract.tex
\documentclass{sigchi-ext}
\usepackage[T1]{fontenc}
\usepackage{textcomp}
\usepackage[scaled=.92]{helvet} 
\usepackage{graphicx} 
\usepackage{balance}  
\usepackage{booktabs} 
\usepackage{ccicons}  
\usepackage{ragged2e} 

\usepackage{array}
\newcolumntype{P}[1]{>{\centering\arraybackslash}p{#1}}
\usepackage{multirow}
\usepackage{arydshln}
\usepackage{enumerate}

\usepackage[utf8]{inputenc} 

\def\plaintitle{Implicit Cooperation: Emotion Detection for Validation and Adaptation of Automated Vehicles' Driving Behavior} 

\def\emptyauthor{}
\def\plainkeywords{Automation Behavior; Emotions; Implicit Interaction; Adaptive Automation; Ambient Intelligence; Safety-Critical Interaction; Automated Vehicles; Human-Machine-Cooperation}

\title{\plaintitle}

\numberofauthors{3}
\author{%
  \alignauthor{%
    \textbf{Henrik Detjen}\\
    \affaddr{Positive Computing Institue} \\
    \affaddr{University of Applied Science Ruhr West} \\
    \affaddr{Essen, DE} \\
    \email{henrik.detjen@hs-ruhrwest.de} } 
    \vfil \alignauthor{%
    \textbf{Stefan Geisler}\\
    \affaddr{Positive Computing Institue} \\
    \affaddr{University of Applied Science Ruhr West} \\
    \affaddr{Bottrop, DE} \\
    \email{stefan.geisler@hs-ruhrwest.de} } 
    \vfil \alignauthor{%
    \textbf{Stefan Schneegass}\\    
    \affaddr{HCI Group} \\
    \affaddr{University of Duisburg-Essen} \\
    \affaddr{Essen, DE}
    \email{stefan.schneegass@uni-due.de} } }

\definecolor{linkColor}{RGB}{6,125,233}
\hypersetup{%
  pdftitle={\plaintitle},
  pdfauthor={\emptyauthor},
  pdfkeywords={\plainkeywords},
  bookmarksnumbered,
  pdfstartview={FitH},
  colorlinks,
  citecolor=black,
  filecolor=black,
  linkcolor=black,
  urlcolor=linkColor,
  breaklinks=true,
}

\begin{document}

\CopyrightYear{2020}
\setcopyright{rightsretained}
\conferenceinfo{CHI'20 Workshop on ``Momentary Emotion Elicitation \& Capture''}{April  25--30, 2020, Honolulu, HI, USA}
\isbn{978-1-4503-6819-3/20/04}
\doi{https://meec-ws.com/}
\copyrightinfo{\acmcopyright}

\maketitle

\RaggedRight{} 

\begin{abstract}
Human emotion detection in automated vehicles helps to improve comfort and safety.
Research in the automotive domain focuses a lot on sensing drivers' drowsiness and aggression. 
We present a new form of implicit driver-vehicle cooperation, where emotion detection is integrated into an automated vehicle's decision-making process.
Constant evaluation of the driver's reaction to vehicle behavior allows us to revise decisions and helps to increase the safety of future automated vehicles.
\end{abstract}

\keywords{\plainkeywords}

\ccsdesc[500]{Human-centered computing~Human computer interaction (HCI)}
\ccsdesc[300]{Human-centered computing~Ubiquitous and mobile computing}

\ccsdesc[300]{Hardware~Sensors and actuators}
\ccsdesc[500]{Human-centered computing~Ubiquitous and mobile devices}

\printccsdesc

\input{content}

\balance{} 

\bibliographystyle{SIGCHI-Reference-Format}
\bibliography{extended-abstract.bib}

\end{document}

%% file: content.tex
\section{Introduction}
When driving automated, sensors might work restricted, e.g., due to a car driving in front. 
In these situations, human intervention is required.
Such interventions might be completely manual, or, to maintain comfort, in a cooperative form: 
The human extends the car's restricted sensor range and resulting insecurities by providing his view and judgment of the situation, e.g., by deciding if overtaking is safe or not~\cite{10.1145/3342197.3344531}.
The car leaves the final decision to take action to the human driver/passenger and executes all other driving-related actions, which keeps the driving comfort for the human.
In this paper, we want to discuss the idea of a human intervention which requires no explicit interaction, but yet achieves safer behavior of an automated vehicle.
In particular, we suggest using implicit emotional states of the driver as an additional input to confirm or cancel the planned vehicle actions.

\section{Cooperative, Implicit Decision Making While Driving}
Typically, cooperative driving means that the human intervenes in the decision-making process of an automated vehicle. This kind of cooperation has the potential to combine the strengths of manual driving with automated cars~\cite{detjen2019maneuver}.
The driver/passenger is asked for a specific input (option or information), for example, if the car should overtake or not, and then the car executes the actions. 
These interventions are commonly performed explicitly by giving voice, gesture~\cite{10.1145/3340764.3340798}, or touch~\cite{5164468} commands.
Implicit interaction modalities have to be interpreted by the system through a set of parameters.
So this kind of interaction does not play an important role, as it might not completely match the user's intention.
When driving on the highway and the car interprets the driver/passenger's unintended lean to the left as an implicit command to perform a lane change to the left, for example, users will get annoyed and mistrust the system.
Hence, implicit interaction seems to be unfeasible for the activation/triggering of driving maneuvers.

In human interaction, when we evaluate ideas together, think about the next meal or speak with another person, the confidence of our thoughts and decisions and our resulting behavior is expressed through multiple, hard to control body parameters.
Most of these behavioral/affective parameters are conveyed on implicit interaction channels.
If we feel secure, we look confident, our thinking is clear, and we are less aroused.
If we feel insecure, we look frightened, we cannot think clearly, and our pulse quickens. 
Humans can easily interpret these channels and tell such secure or insecure behavior apart.
So what if a machine could do, too.
Emotion detection can be used to enable a affect-sensitive human-robot cooperation~\cite{rani_sarkar_smith_kirby_2004}.

Even if the driver/passenger does not decide actively to take a particular action during autonomous driving modes, but the car does, he or she is affected by the machine's actions.
If (s)he is comfortable with the car's driving, (s)he is relaxed, but an aggressive driving style leads to fearful reactions.
So, the passenger's behavioral/affective state reflects the level of driving comfort and safety.

\begin{figure*}
  \centering
  \includegraphics[width=\textwidth]{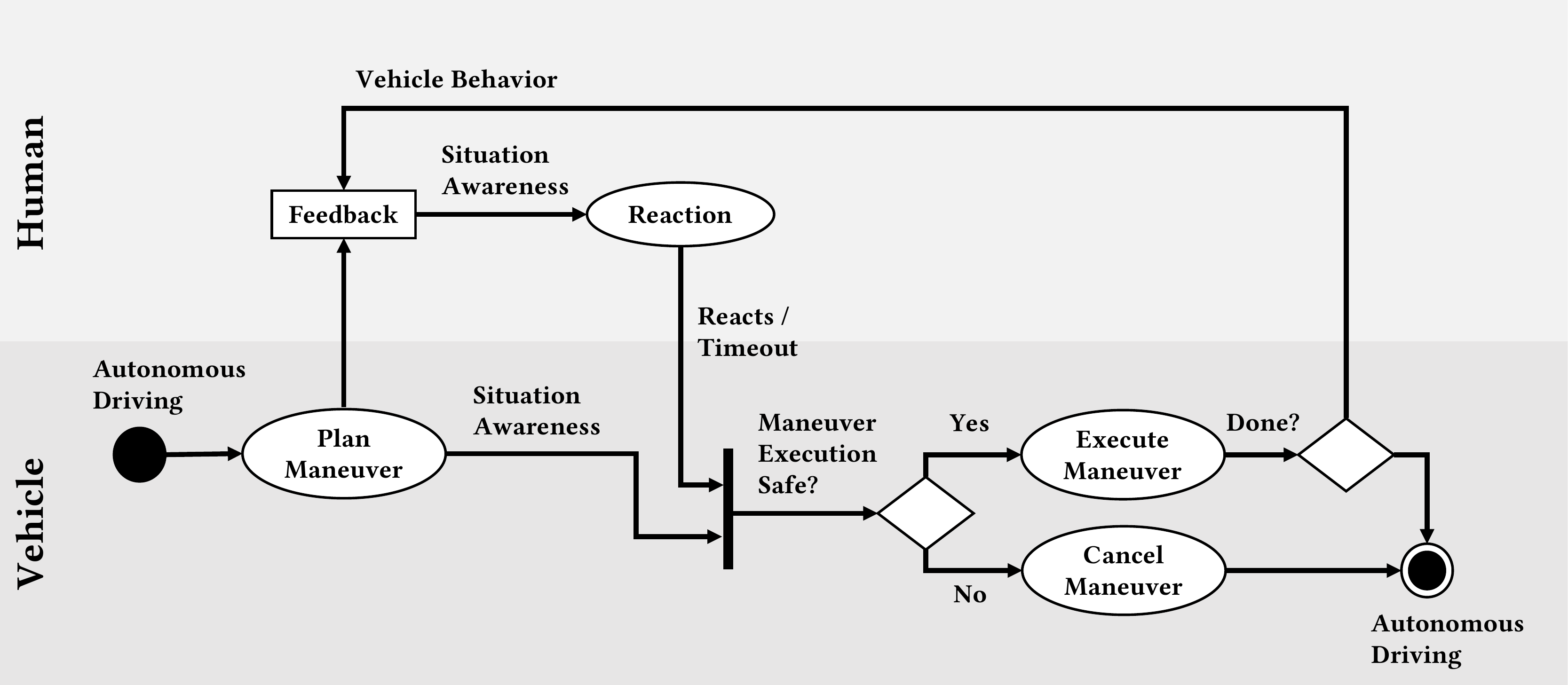}
  \caption{Cooperative Maneuver Execution Process with Constant Evaluation of Vehicle Decisions through Implicit Driver Feedback}
    \label{fig:bla}
\end{figure*}

In consequence, if a machine could interpret the affective state of humans, it could also interpret the level of safety and comfort their actions induce. 
We suggest using this human-confidence-parameters to support the vehicle's decision-making process as well as the decision rollback process.
In this paper, we want to analyze the validation and rollback of decisions.
That requires a form of human-machine-cooperation, where one can cancel the execution of actions.
It requires a continuous decision evaluation process during the execution phase of a maneuver.
Further, decisions have to be evaluated with human affects/behavior and measured in real-time.
The car could implement a loop which constantly evaluates a maneuver execution with the sensed confidence through its sensors and human reaction against a safe threshold:
$Conf(M)_{SAFE} >= Conf(M)_{OWN} - Conf(M)_{PASSENGERS}$.
Figure~\ref{fig:bla} summarizes the proposed cooperation process visually.

\subsection{An Exemplary Scenario: Canceled Overtaking}
It is dark and rainy on the highway.
You drive home from work with your highly automated car.
While driving in the autonomous mode, you are having a warm meal.
Your car decides to perform a lane change to overtake a slower car ahead.
In some distance behind your car, headlights of another car are approaching.
Your car informs you about the planned action on a head-up display and though a special announcement sound.
You stay calm because you have experienced this situation many times before. 
Then, the car performs the lane change.
Due to the rain, the radar and ultrasonic sensors of your car get a little bit distorted.
Hence, the calculated level of the system's confidence to perform the lane change is just over the safe threshold. 
There is some probability that the confidence is unjustified.
Unfortunately, it is.
The car behind is much faster than expected.
You get nervous as the car starts to perform the lane change because you cannot reach the steering wheel timely without spilling your hot coffee.
The approaching car in only a few meters away.
Luckily, your car's driver emotion-sensing system has noticed your anxiety.
It reevaluates its lane change decision and sets down the calculated confidence level for a few percent.
The new confidence value is below the safe threshold, and the car cancels the maneuver just in time.

\subsection{Further (Non-)Use Cases}
For safety-related driving assistants, the inclusion of human reactions can have a negative effect. 
When the car performs an emergency brake, the action should not be interrupted through humans feeling uncomfortable, or they would not be safe anymore.
However, for other use cases, the validation of car actions through human emotions seems useful. Some examples: adjusting speed, adjusting accelerating/breaking behavior, adjusting the distance, changing lanes, turning, automated parking\dots

\setlength{\dashlinedash}{0.5pt}
\setlength{\dashlinegap}{2.5pt}
\setlength{\arrayrulewidth}{0.5pt}
\begin{table*}[t!]
  \centering
  \resizebox{\textwidth}{!}{
  \begin{tabular}{l l l}
    \toprule
    \textbf{Body Parameter}
    & \textbf{Required Technology}
    & \textbf{Study Example} \\
    \midrule
    Facial expressions & Video camera & Facial expressions analysis to detect drowsiness~\cite{10.1007/978-3-540-75773-3_2} \\
    \hdashline
    Pupil diameter & Eye-tracking camera / glasses & Automatic stress classification in the car~\cite{doi:10.1080/10447318.2013.848320} \\
    \hdashline
    Gaze & Eye-tracking camera / glasses & Pre-crash gaze behavior to predict crash intensity~\cite{SEPPELT201748} \\
    \hdashline
    Voice & Microphone & Language reliability display to improve UX~\cite{10.1145/3349263.3351312}, \\
     &  & detecting emotions through voice parameters~\cite{jones2005automatic} \\
    \hdashline
    Gestures & Video camera & -  \\
    \hdashline
    Body position & Video camera, & Postures to detect driver activities~\cite{mci/Riener2007} \\
    & In-seat force sensors &  \\
    \hdashline
    Brain activity & EEG & Adapt lights in the car to driver arousal~\cite{hassib2019detecting} \\
    \hdashline
    Heart rate & Pulse meter, ECG & Stress correlates with heart rate~\cite{1438384}, HRV-analysis~\cite{10.1145/1620509.1620529} \\
    \hdashline
    Electro-dermal activity & EDA-sensor & Stress correlates with skin conductance~\cite{1438384} \\
    \hdashline
    Cortisol level & Magnetic resonance imaging, & - \\
    &  computerized tomography &  \\
    \hdashline
    Thermal response & Thermal camera & Driver's emotion recognition through thermal imaging~\cite{6024802} \\
    \bottomrule
  \end{tabular}
  }
  \caption{Body parameters responding to human emotions and how to measure them in the vehicle}~\label{tab:table1}
\end{table*}

\section{Sensing Human Emotions in the Car}
For our scenario, two emotions are of special interest: surprise and fear. 
Both are linked to a rather negative valence and high arousal.
However, driver emotion detection applications have focused a lot on drowsiness and high arousal/load detection.
In the following, we depict some examples.

Völkel et al.~\cite{10.1145/3239092.3267102} test two app concepts which utilize the driver's state. First, a dashboard app showing safety-critical states like drowsiness or aggressiveness and second, a warning app that gave feedback to the driver when such an emotion reached a certain threshold. Participants preferred to receive only safety-critical notifications (high threshold).
A technical implementation such a system is planned by Vasey, Ko, and Jeon~\cite{10.1145/3239092.3267417}.
For a complete review of drowsiness detection, we refer to Sahayadhas, Sundaraj, and Murugappan~\cite{Sahayadhas_2012}.
Healey, Theocharous and Kveton~\cite{Healey2010} investigate the reactions of passengers on an aggressive driver's driving behavior. They found that the fear of passengers correlates highly with galvanic skin response. They formulate the idea to report the passenger's condition back to the driver, yet they did not test it. 
In contrast to existing work, we do not confront the driver/passenger with the vehicle behavior after a situation (e.g., \cite{WOUTERS2000643}), but want to access their passive reactions in real-time and integrate them in the vehicle's decision evaluation.

\subsection{Sensing Channels}
The body emits emotions on multiple channels.
Via affective computing, we can try to approximate human emotion through measurable channels~\cite{10.1109/T-AFFC.2010.1}.
Not all of these channels are suited to sense human emotions in the car.
In the following, we will discuss different channels, technologies, and their potential for practical application in automated vehicles in the near future (see Table~\ref{tab:table1}).

A downside of all emotion-sensing approaches is that one cannot link the emotional reactions of humans in the car explicitly to the system's driving behavior.
Other environmental factors, like talking to another passenger or watching a horror movie, can lead to arousing emotional states.
That is a clear limitation, but we also suggest that future vehicles that already have complex real-time emotion detection, will also have passenger activity detection. 
Thus, we expect the car to tell these differences apart by using other contextual parameters.
Some of the emotion-sensing channels are also well suited to detect driver/passenger activities, which leads to synergistic effects.
One should consider such synergies for the rating of future sensing methods in the car.

\subsection{Discussion of Technologies}
Unobtrusive methods like facial expression, gaze, pupil diameter, and thermal imaging seem feasible for the use in automated vehicles in the near future because it is not likely that users put effort into equipping or install a special device before each ride. 
Further, these channels are constantly available, in contrast to speech, for example.
Cameras are already used in vehicles for driver state detection, primary driver drowsiness detection nowadays.
Therefore, camera-based methods are the predestined source for driver emotion detection.
Currently, cameras are commonly installed in the front only. That makes the detection impossible when the is facing in the ``wrong'' direction.
Nevertheless, future vehicles might have more cameras installed in the cabin to enable complete detection without gaps.
Many drivers/passengers also wear (sun-)glasses, which harden or hinder the detection, depending on the method.
Hence, further channels might be necessary to ensure steady detection.
Additional input sources become available when drivers/passengers wear devices like smartwatches.
HRV- or skin conductance analysis is possible with pulse sensors.
Car and smartwatch/app companies should work on common software interfaces to realize this.

\section{Conclusion \& Future Work}
In the current state, there are major weaknesses in the practical use of each emotion detection technology.
Thus, emotion detection in the car has to follow a multi-method paradigm to overcome these weaknesses and dynamically adapt to a different channel and new technological developments (Plug \& Play Sensors).
In the automotive domain, emotion-sensing focuses on detecting drowsiness or aggression. 
The system uses the sensed emotions to reflect the emotional state to the driver.
We present a use case where implicit driver reactions are integrated into the car's decision evaluation process.
In our future research, we will elicit and analyze a data set with human emotional reactions on the near and full failure of automated systems.
For our presented scenario, we will also investigate how emotion detection might help to prevent or weaken the impact of accidents.